\documentclass{elsart}
\usepackage{epsfig}
\usepackage{amssymb}
\journal{Nuclear Physics A}
\def\v#1{\mbox{\boldmath$#1$}}
\def\ket#1{|#1 \rangle}
\def\bra#1{\langle #1|}
\begin{document}
\begin{frontmatter}
\title{Ground state normalization in the nonmesonic weak decay
of $^{12}_{\Lambda}C$ hypernucleus within a nuclear matter formalism.}

\author{E. Bauer}

\address{
Departamento de F\'{\i}sica, Universidad Nacional de
La Plata,\\
C. C. 67, 1900 La Plata, Argentina}
\address{
Instituto de F\'{\i}sica La Plata,
CONICET, 1900 La Plata, Argentina}

\begin{abstract}
The nonmesonic weak decay width of $^{12}_{\Lambda}C$ hypernucleus
has been evaluated within a nuclear matter formalism, using the
local density approximation. In addition to the one-body induced
decay $(\Lambda N \rightarrow n N)$, it has been also considered
the two-body induced decay $(\Lambda NN \rightarrow n N N)$. This
second decay is originated from ground state correlations, where a
renormalization procedure to ensure a ground state normalized to
one has been implemented. Our results show that the plain addition
of the two-body induced decay implies a lost in the ground
state-norm, which adds $\sim 38\%$ of spurious intensity to the
nonmesonic weak decay width.
By an adequate selection of the $\Lambda N$-transition potential,
our result for the
nonmesonic weak decay width of $^{12}_{\Lambda}C$ is 0.956, in good
agreement with the most recent data.
\end{abstract}
\end{frontmatter}

\newpage
\section{Introduction}

\label{intro}
A $\Lambda$-hypernucleus decays via the weak interaction mainly by
two decay mechanisms: the so-called mesonic decay $(\Lambda
\rightarrow \pi N)$ and the nonmesonic one $(NM)$, where no meson
is present in the final state (for review articles see
\cite{ra98,al02}). The $NM$-decay width is denoted as
$\Gamma_{NM}$, which is defined as the sum of $\Gamma_{1} \equiv
\Gamma(\Lambda N \rightarrow n N)$ plus $\Gamma_{2} \equiv
\Gamma(\Lambda NN \rightarrow n N N)$. The $\Gamma_{1}$-decay
width itself is the sum of $\Gamma_{n} \equiv \Gamma(\Lambda n
\rightarrow n n)$ plus $\Gamma_{p} \equiv \Gamma(\Lambda p
\rightarrow n p)$. Experimental values are given for
$\Gamma_{NM}$, the ratio $\Gamma_{n/p} \equiv
\Gamma_{n}/\Gamma_{p}$ and the asymmetry of the protons emitted in
the $NM$ decay of polarized hypernuclei. In the present
contribution, we focuss on $\Gamma_{NM}$, evaluated in nuclear
matter together with the local density approximation which allows
us to analyze the $^{12}_{\Lambda}C$ hypernucleus.

In the past, it has been an usual statement to assert that while
the theory accounts for the experimental values of $\Gamma_{NM}$,
the same is not true for the ratio $\Gamma_{n/p}$. In fact, the
disagreement between the theoretical and the experimental value for
this ratio, has been named as "the $\Gamma_{n/p}$-puzzle". This
situation has changed in recent years: new theoretical analysis
together with more experimental information, have led us to a solution of
the so-called puzzle. A typical theoretical value for the ratio for $^{12}_{\Lambda}C$
is $\Gamma_{n/p} \sim 0.3$, while data analyzed by means of the
intranuclear cascade code (INC)~\cite{ra97,ga03,ga04,ba06}, gives a
result $\Gamma_{n/p}^{exp} \sim 0.4 \pm 0.1$\footnotemark
{\footnotetext{For this result, it has been considered the
$\cos(\theta)< -0.80$ region (where this angle is the
one between the two outgoing nucleons)
and a kinetic detection threshold
for nucleons $T^{th}_{N}=30$~MeV.}}. However,
it should be noted that there still exist discrepancies with some
nucleon spectra. For instance, the experimental single coincidence
proton spectra for $^{12}_{\Lambda}C$ is not well reproduced yet.

In nuclear matter (using the local density approximation), some reported
calculations for $\Gamma_{1}$ for $^{12}_{\Lambda}C$, have values in the range
between 0.5~\cite{du96} up to 1.45~\cite{os85} (given in units of
the $\Lambda$-free decay width, $\Gamma^{0}$). While typically
$\Gamma_{2}/\Gamma_{1} \sim 0.3$. The most recent experimental
determination of $\Gamma_{NM}$ has been done by Outa et al.~\cite{ou05},
whom have reported a value $\Gamma_{NM}^{exp}=0.940 \pm 0.035$.
Some previous experimental determinations are
$\Gamma_{NM}^{exp}=1.14 \pm 0.20$~\cite{Sz91},
$\Gamma_{NM}^{exp}=0.89 \pm 0.15 \pm 0.03$~\cite{No95} and
$\Gamma_{NM}^{exp}=0.828 \pm 0.056 \pm 0.066$~\cite{sa05}.
There is some incompatibility between the result in~\cite{sa05}, due
to Sato et al. and both Noumi et al.~\cite{No95} and Outa et al.
We have relied on the Outa result, not only because it is the most recent one,
but also due to it compatibility with
both~\cite{Sz91} and \cite{No95}-values. Beyond this controversy,
the more precise determination of $\Gamma_{NM}^{exp}$ offers us the
opportunity to revise the theoretical determination of
$\Gamma_{NM}$. In two previous works, a model for the
evaluation of $\Gamma_{1}$ and $\Gamma_{2}$ have been
developed (see~\cite{ba03} and \cite{ba04}, respectively). Our
scheme employs the same microscopic formalism and interactions
for both $\Gamma_{1}$ and $\Gamma_{2}$. Within this model,
the reproduction of $\Gamma_{NM}^{exp}$
seems not possible: the predicted values are always too big.
The main concern of this work is to understand and solve this problem.

To solve this problem it has been revised the
way in which $\Gamma_{2}$ is added
to $\Gamma_{1}$, to build up $\Gamma_{NM}$.
The $\Gamma_{2}$-contribution is originated from
ground state correlations and the simple addition of $\Gamma_{2}$
plus $\Gamma_{1}$, would add some spurious intensity because the
ground state is not normalized to one. This point turns out to
be relevant not only in the determination of $\Gamma_{NM}$,
but also for the $\Gamma_{2}/\Gamma_{1}$-ratio, which is used as an
input in the determination of $\Gamma_{n/p}^{exp}$. As a
further comment on this point, the lack in the normalization
is not restricted to our particular model, but to any calculation
where $\Gamma_{2}$ is considered. A second point
refers to the implementation of short range correlations ($SRC$)
in nuclear matter. It is shown that for $\Gamma_{2}$ this has to be done with
some particular care due to numerical reasons.
The implementation of these two points in
the already developed formalism for $\Gamma_{1}$ and $\Gamma_{2}$,
gives a $\Gamma_{NM}$-value in good agreement with data.

This work is organized as follows. In Sect.~\ref{gsc}, our model for
the renormalization of the ground state is presented, showing a
scheme to add $\Gamma_{2}$ to $\Gamma_{NM}$. In Sect.~\ref{src},
a model for the $SRC$ in nuclear matter is discussed in detail.
Numerical results are shown in Sect.~\ref{results}, together with
an analysis of the implications of the corrections. Finally,
in Sect.~\ref{conclusions}, some conclusions are given.

\section{Ground state correlations ($GSC$)}

\label{gsc}
To start with, we write down the partial decay width
$\Gamma_{NM}(k_{F})$ in a schematic way as,
\begin{equation}
\label{decw} \Gamma_{NM}(k_{F}) = \sum_{f} \,
 |\bra{f} V^{\Lambda N} \ket{0}_{k_{F}}|^{2}
\delta^{(4)}(p_{f}-p_{0}),
\end{equation}
where $\ket{0}_{k_{F}}$ and $\ket{f}$ are the ground state and the
final state, respectively; $V^{\Lambda N}$ is the two-body
transition potential and $p_{i}$ represents an energy-momentum
four-vector. The Fermi
momentum is denoted as $k_{F}$. By performing the integration over
$k_{F}$ using the local density approximation
(see~\cite{os85}), the total decay width $\Gamma_{NM}$ is
obtained. Now, the ground state can be written as,
\begin{equation}
\label{gstate} \ket{0}_{k_{F}}={\cal N}(k_{F}) \, \left(\ket{\;}_{k_{F}}
- \frac{1}{4} \, \sum_{p_{1}, p_{2}, h_{1}, h_{2}} \,
\frac{\bra{p_{1} p_{2} h_{1} h_{2}} V^{N N} \ket{\;}_{k_{F}}}
{\epsilon_{p1}+\epsilon_{p2}-\epsilon_{h1}-\epsilon_{h2}} \,
\ket{p_{1} p_{2} h_{1} h_{2}}\right),
\end{equation}
where the second term in the right hand side of the equation
represents $2p2h$-correlations. In this equation
$\ket{\;}_{k_{F}}$, is the Hartree-Fock vacuum. In the
denominator, $\epsilon_{i}$ are the single particle energies. The
nuclear residual interaction is represented by $V^{N N}$ and
${\cal N}(k_{F})$ is the normalization as a function of $k_{F}$:
\begin{equation}
\label{norconst} {\cal N}(k_{F})=\left( 1 + \frac{1}{16} \, \sum_{p_{1},
p_{2}, h_{1}, h_{2}} \, \left|\frac{\bra{p_{1} p_{2} h_{1} h_{2}}
V^{N N} \ket{\;}_{k_{F}}}
{\epsilon_{p1}+\epsilon_{p2}-\epsilon_{h1}-\epsilon_{h2}}\right|^{2}
\, \right)^{-1/2}.
\end{equation}
We will show soon that the inclusion of ${\cal N}(k_{F})$, has an
important effect over $\Gamma_{NM}$. The importance of a proper
treatment of the ground state normalization has been already
pointed out by Van Neck et al.~\cite{va92}. When
Eq.~(\ref{gstate}) is inserted into the expression of
$\Gamma_{NM}(k_{F})$ given by Eq.~(\ref{decw}) with the arbitrary
selection of ${\cal N}(k_{F})=1$, the usual expressions for $\Gamma_{1}$
and $\Gamma_{2}$, are obtained. The first one comes from the first
term in Eq.~(\ref{gstate}), while $\Gamma_{2}$ results from the
second term in the same equation.

Alternatively, if the $GSC$ are neglected (\textit{i.e.}
$\Gamma_{2}=0$), then $\ket{0}_{k_{F}}=\ket{\;}_{k_{F}}$ and
${\cal N}(k_{F})=1$. However, when $GSC$ are included, the use of
${\cal N}(k_{F})=1$ means that the ground state is not properly
normalized and therefore some spurious intensity is added
to $\Gamma_{NM}$.

\section{Short range correlations ($SRC$)}

\label{src}
In momentum space one model to take care
of $SRC$ is by the use of a modified transition potential obtained as,
(see~\cite{os82}),
\begin{equation}
\label{srcp} V_{SRC}(\v{q}) \, = \, V(\v{q}) \, - \, \int \frac{d
\v{p}}{(2 \pi)^3} \tilde{\xi}(|\v{p} + \v{q}|) \, V(\v{p}),
\end{equation}
where we employ,
\begin{equation}
\label{cfunp} \tilde{\xi}(p) = \frac{2 \pi^2}{q_c^2} \,
\delta(p-q_c),
\end{equation}
with $q_c =780$ MeV/c, as a particular correlation function
in momentum space. We have limited our discussion of $SRC$
to this model and it implementation in the evaluation of $\Gamma_{NM}$
deserves some care. We show this with an example. Let us show the result of
Eq.~(\ref{srcp}) with the central part of the parity conserving
one pion exchange potential, which we write in a simplified manner
as,
\begin{equation}
\label{pionc} V_{\pi}^C(\v{q})  =  C_{\pi} \;
\frac{\v{q}^2}{\v{q}^2+m_{\pi}^2} \v{\sigma_1} \cdot \v{\sigma_2}
\; \v{\tau_1} \cdot \v{\tau_2}
\end{equation}
with $C_{\pi} = - G_F m_{\pi}^2 \; (g_{NN \pi}/2 M) \; (B_{\pi}/2
\bar{M})$, where $\bar{M}$, is the average between the nucleon and
$\Lambda$ masses. Using this potential in Eq.~(\ref{srcp}) we
obtain,
\begin{eqnarray}
\label{srcpc1} V_{\pi}^{SRC, \, C}(\v{q}) &  & = \,
V_{\pi}^C(\v{q})
\, -  \nonumber \\
&  & C_{\pi} \frac{1}{2} \{ 2 + \frac{m_{\pi}^2}{2 q_c |\v{q}|}
\ln |\frac{q_c^2 + m_{\pi}^2 + \v{q}^2 - 2 q_c |\v{q}|}
      {q_c^2 + m_{\pi}^2 + \v{q}^2 + 2 q_c |\v{q}|}| \}
\v{\sigma_1} \cdot \v{\sigma_2} \; \v{\tau_1} \cdot \v{\tau_2}
\end{eqnarray}
by calling $\kappa = 2 q_c |\v{q}|/(q_c^2 + m_{\pi}^2 + \v{q}^2)$
and making the approximation,
\begin{equation}
\label{aproxlog} \ln (1 + \kappa) \approx \kappa
\end{equation}
we finally obtain,
\begin{equation}
\label{srcpc2} V_{\pi}^{SRC, \, C}(\v{q}) \, = \, V_{\pi}^C(\v{q})
\, - \, C_{\pi} \frac{q_c^2 + \v{q}^2}{q_c^2 + m_{\pi}^2 +
\v{q}^2} \v{\sigma_1} \cdot \v{\sigma_2} \; \v{\tau_1} \cdot
\v{\tau_2}
\end{equation}
which is equivalent to the following general prescription to build up the
modified potential due to the action of the $SRC$:
\begin{equation}
\label{srcpc3} V^{SRC, \, C}(\v{q}) \, = \, V^C(\v{q})
\, - V^C(\v{q}^2 \rightarrow q_c^2 + \v{q}^2).
\end{equation}
To the best of our knowledge, this way of taking care of $SRC$ in
nuclear matter is the most frequently used one.
However, we should call attention on the non-equivalence between
the approximation in Eq.~(\ref{srcpc3}) and the one in Eq.~(\ref{srcp})
for the kinematical conditions of the nonmesonic $\Lambda$-decay.
This is because the approximation
given by Eq.~(\ref{aproxlog}), is a bad approximation for the
momentum transfer in the $\Lambda N \rightarrow NN$
decay channel (where $q \approx 400$ MeV/c). A
simple numerical test shows that Eq.~(\ref{srcpc3}) fairly
accounts for the expression given by Eq.~(\ref{srcp}) only for $q
\lesssim 50$ MeV/c.
For the full $V^{\Lambda N}$-transition potential (which includes
$q$-dependent form factors), the integral in Eq.~(\ref{srcp}) can
be also performed analytically. The numerical results show that
$\Gamma_{NM}$ evaluated with the inclusion of $SRC$ given by
the model in Eqs.~(\ref{srcp}) and (\ref{cfunp}),
is $\sim 35 \%$ smaller than the same quantity
with the prescription in Eq.~(\ref{srcpc3})
(employing the same $q_c$-value).
In the present contribution we present results only for the model
in Eqs.~(\ref{srcp}) and (\ref{cfunp}), as
this model gives us some confidence about
it applicability within a wide range in the variation
of the momentum transfer.

Due to it frequent use, it is important
to discuss the prescription in Eq.~(\ref{srcpc3}), which
is in fact, an approximation to the model in
Eqs.~(\ref{srcp}) and (\ref{cfunp}).
The employment of this prescription would be particularly
questionable in the evaluation of $\Gamma_{2}$
(rather than $\Gamma_{1}$), for the
reasons that follows. Let us write down
explicit expressions for both $\Gamma_{1}$ and $\Gamma_{2}$. We do
this in a very schematic way,
\begin{equation}
\label{gam1p1h}
\Gamma_{1}(\v{k},k_{F})   =  C
\int d  \v{q}
\theta(q_{0})  \theta(|\v{k}-\v{q}| - k_F)
(V^{\Lambda N}(q))^{SRC}
Im \Pi_{1p1p}(q_0,\v{q})
\end{equation}
and
\begin{equation}
\label{gam2p2h}
\Gamma_{2}(\v{k},k_{F})   =  C
\int d  \v{q}
\theta(q_{0})  \theta(|\v{k}-\v{q}| - k_F)
(V^{\Lambda N}(q))^{SRC}
Im \Pi_{2p2p}(q_0,\v{q})
\end{equation}
where $C=- 6 (G_F m_{\pi}^2)^{2} \pi/(2 \pi)^{3}$ and
$q_0=k_0-E(\v{k}-\v{q})-V_N$, being $k$ the energy-momentum
of the $\Lambda$. Final values for $\Gamma_{1}$ and $\Gamma_{2}$
are obtained after integrating over $\v{k}$ and $k_F$. The
functions $\Pi_{1p1p}$ and $\Pi_{2p2p}$ are the $1p1h$ and
$2p2h$-polarizations functions, respectively. We do not go
through the derivation of these expressions (details can be
found in~\cite{ra94}, for instance).
\begin{figure}[h]
\centerline{\includegraphics[scale=0.57]{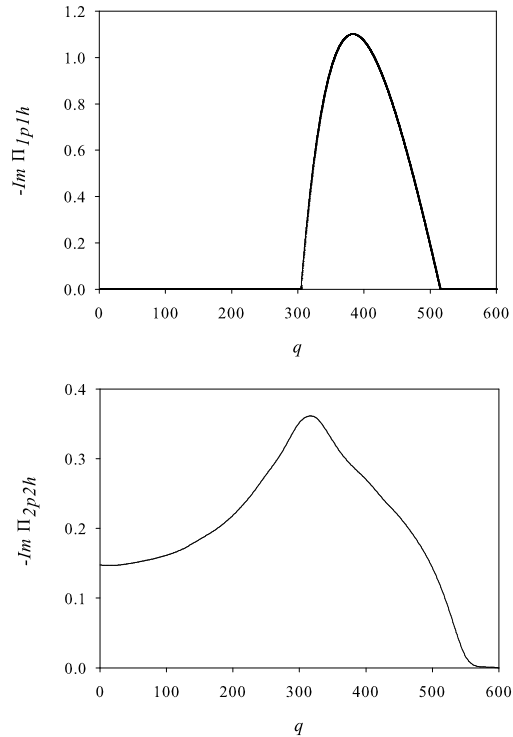}}
\caption{Imaginary part of the polarization functions $\Pi_{1p1h}$ and
$\Pi_{2p2h}$ as a function of the momentum transfer $q$.
The polarization functions and $q$, are in units of
$10^{-3}$~MeV$^{-1}$~fm$^{-3}$ and MeV/c, respectively.}
\label{fig1}
\end{figure}
In Fig.~\ref{fig1}, we
have plotted $Im \Pi_{1p1p}$ and $Im \Pi_{2p2p}$ as a function
of the momentum transfer $q$. For simplicity, this is done
for a $\Lambda$ at rest ($\v{k}=0$) and for $k_F=210$~MeV/c.
The behavior of $Im \Pi_{1p1p}$, is a narrow function,
peaked around $q \approx 400$ MeV/c. This range of variation in
$q$, makes the approximation in Eq.~(\ref{srcpc3}) acceptable once
$q_c$ is somehow adjusted. Let us be clear: with the same $q_c$-value, the
results from Eq.~(\ref{srcp}) and Eq.~(\ref{srcpc3}) are
different because $Im \Pi_{1p1p}$ is non-zero far way from
$q \lesssim 50$ MeV/c. But the narrow $q$-variation
establish by $Im \Pi_{1p1p}$, would make the
approximation in Eq.~(\ref{srcpc3}) acceptable, once
the $q_c$-value is adjusted using some observable or
by a comparison with a finite nucleus calculation.

From the same figure, the situation for $Im \Pi_{2p2p}$ is
very different as it is spread over a wide $q$-region.
The use of Eq.~(\ref{srcpc3}) would introduce a systematic
error in the evaluation of $\Gamma_{2}$, because due
to a numerically wrong approximation, the $SRC$ are
incorrectly weighed for different $q$-values.
This warning is not restricted to our particular
$\Gamma_{2}$-evaluation. There are two former models
for the evaluation of $\Gamma_{2}$. The starting point of
all this evaluations (together with the one of ours) is
the Eq.~(\ref{gam2p2h}), but they differ between each other
in the model for $\Pi_{2p2h}$.
The first work which has called attention on
$\Gamma_{2}$ is the one due to Alberico {\it et al.} \cite{al91},
where a so-called semi-phenomenological $\Gamma_{2}$ has been
adopted, which results from a microscopic evaluation of the
polarization propagator $\Pi_{2p2h}$ in nuclear matter, originally
performed for electron scattering in~\cite{al84}. Using this
electron scattering calculation, a constant $Im \Pi_{2p2h}$
is proposed, which is appropriate for pion absorption.
Thereafter, Ramos {\it et al.}~\cite{ra94}, has used also a
semi-phenomenological $\Gamma_{2}$, where
an approximate value for the function $Im \Pi_{2p2h}$, is
obtained as the product of the phase space corresponding to the
$\Lambda NN \rightarrow NNN$-reaction, times
a constant taken from pion absorption.
It should be noted that in the $\Lambda NN \rightarrow NNN$-reaction,
all mesons are strictly off the mass shell and the
employment of the pion absorption results are used as
an approximation to take care of the dynamics involved in
the evaluation of $\Gamma_{2}$.
Beyond the same starting point of
Eq.~(\ref{gam2p2h}) and the difference in the
calculation of the function $Im \Pi_{2p2h}$, these two works
also differ from the one in~\cite{ba04}, by
the way in which the isospin is taken into consideration and
some minor points. In any case, the just quoted warning
in the inclusion of the $SRC$ is valid for all these
$\Gamma_{2}$-models.

\section{Results and discussion}

\label{results}
We turn now to the numerical results. The transition potential
$V^{\Lambda N}$, is represented by the exchanges of the $\pi$,
$\eta$, $K$, $\rho$, $\omega$ and $K^*$-mesons, whose formulation
has been taken from \cite{pa97}, and the values of the different
coupling constants and cutoff parameters appearing in the
transition potential have been taken from \cite{na77} and
\cite{st99}, named as Nijimegen and NSC97f, respectively. For the
nuclear residual interaction $V^{N N}$ (which is employed in both
$\Gamma_{2}$ and ${\cal N}(k_{F})$), we have used the Bonn
potential~\cite{ma87} in the framework of the parametrization
presented in~\cite{br96}, which contains the exchange of $\pi$,
$\rho$, $\sigma$ and $\omega$ mesons, while the $\eta$ and
$\delta$-mesons are neglected. In implementing the LDA, the
hyperon is assumed to be in the $1s_{1/2}$ orbit of  a harmonic
oscillator well with frequency $\hbar \omega = 10.8$ MeV, where we
have employed different values for the proton and neutron Fermi
momenta, $k_{F_n}$ and $k_{F_p}$, respectively (for details
see~\cite{ba07b}).

The partial decay widths $\Gamma_{1}(k_{F})$ and
$\Gamma_{2}(k_{F})$ have been evaluated
using the scheme developed in~\cite{ba03} and \cite{ba04},
respectively; but with the implementation of the
$SRC$ described above. The $\Gamma_{2}$-contribution is built up
to three terms: $\Gamma_{2}=\Gamma_{nn}+\Gamma_{np}+\Gamma_{pp}$,
with $\Gamma_{nn} \equiv \Gamma(\Lambda nn \rightarrow nnn)$,
$\Gamma_{np} \equiv \Gamma(\Lambda np \rightarrow nnp)$ and
$\Gamma_{pp} \equiv \Gamma(\Lambda pp \rightarrow npp)$. The
dominant term is $\Gamma_{np}$, where the relative magnitude of
each contribution follows approximately the relation,
$\Gamma_{np}:\Gamma_{pp}:\Gamma_{nn} \approx 0.78:0.17:0.05$. It
should be noted that once the $GSC$ are considered, the partial
decay widths $\Gamma_{1}(k_{F})$ and $\Gamma_{2}(k_{F})$ are multiplied by the
function ${\cal N}(k_{F})$ and then, the $k_{F}$-integration gives the
final $\Gamma_{NM}$. Therefore, the action of the ground state
renormalization is not the plain multiplication of $\Gamma_{1, \,
2}$ by a constant. In this procedure, we have employed the same
nuclear matter model for $\Gamma_{1, \, 2}(k_{F})$ and
${\cal N}(k_{F})$, using the same nuclear residual interaction,
transition potential and the $SRC$-model.

In Table~\ref{gam12}, we present our values for $\Gamma_{1, \ 2}$
and $\Gamma_{NM}$ for the two mentioned sets of transition potential
parameters, with and without the action of ${\cal N}(k_{F})$. In first
place, it is clear that the effect of the ground state
renormalization is important: for both interactions, the spurious
part in $\Gamma_{NM}$ (\textit{i.e.} $100 \times
|\Gamma_{NM}(without \; renorm.)- \Gamma_{NM}(with \;
renorm.)|/\Gamma_{NM}(with \; renorm.)$), is $\sim38\%$. At
variance, the $\Gamma_{1}$ without renormalization does not differ
very much from $\Gamma_{NM}$ ($=\Gamma_{1}+\Gamma_{2}$) with
renormalization. Our final $\Gamma_{NM}$ with renormalization
shows a small decrease (increase) with respect to
$\Gamma_{1}$ without renormalization, for the interaction Nijimegen
(NSC97f). In fact, while for Nijimegen the value
for $\Gamma_{1}$ is greater than the same one for NSC97f; just the
opposite occurs for $\Gamma_{2}$.
This is a consequence of the different weight of
each spin-isospin component in each interaction, together
with the different structure in the
spin-isospin sums between $\Gamma_{1}$ and $\Gamma_{2}$. As a
further point for this paragraph, our $\Gamma_{NM}$-result for
the interaction Nijimegen is in close agreement with the data
from~\cite{ou05}. And so does the result for the interaction NSC97f
with the data from~\cite{sa05}. Although both data have been
included in the table, for the reasons already discussed we
rely on the~\cite{ou05} data and therefore, we consider the
$\Gamma_{NM}=0.956$ value as our final result. Consequently,
the $\Gamma_{2}/\Gamma_{1}$-ratio takes a value 0.28.
It should be stressed that the
$\Gamma_{2}$-contribution represents $22-30\%$ of $\Gamma_{NM}$ and
while there are many theoretical works which deals with the
evaluation of $\Gamma_{1}$, the same does not occur for
$\Gamma_{2}$.
\begin{table}[h]
\begin{center}
\caption{The nonmesonic weak decay width of $^{12}_{\Lambda}C$.
The first column represents the $V^{\Lambda N}$-transition
potential and the second one refers to the inclusion or not of the
normalization factor in the ground state. All $\Gamma$'s are in
units of free $\Lambda$-decay rate, $\Gamma^0= 2.52 \cdot 10^{-6}$
eV.} \vspace{1cm} \label{gam12}
\begin{tabular}{ccccc}   \hline\hline
$model \; int.$ & $renorm.$ &${\Gamma}_{1}$ &
 ${\Gamma}_{2}$&
${\Gamma}_{NM}$
\\  \hline
$\mbox{Nijimegen} \,$ \cite{na77}   & $no$ & 1.031 & 0.289 & 1.320   \\
$\mbox{NSC97f} \,$ \cite{st99}  & $no$ & 0.814 & 0.348 & 1.162  \\
$\mbox{Nijimegen} \,$ \cite{na77}   & $yes$ & 0.747 & 0.209 &  0.956  \\
$\mbox{NSC97f} \,$ \cite{st99}  & $yes$ & 0.590 & 0.250 & 0.840   \\
$\mbox{experiment} \,$ \cite{ou05}  &  &  &  &  $0.940 \pm 0.035$  \\
$\mbox{experiment} \,$ \cite{sa05}  &  &  &  &  $0.828 \pm 0.056 \pm 0.066$  \\
\hline\hline \\
\end{tabular}
\end{center}
\end{table}

In Table~\ref{gamnp}, a similar analysis to the one in
Table~\ref{gam12}, is done for ${\Gamma}_{n, \, p}$ and the ratio
${\Gamma}_{n/p}$, where the theoretical values
are obtained with the scheme in~\cite{ba03}, (but
using the oscillator frequency $\hbar \omega = 10.8$ MeV,
just mentioned). The decay widths ${\Gamma}_{n, \, p, \, nn \, ...}$
are primary decays. This means that to extract
the ratio ${\Gamma}^{exp}_{n/p}$, from the experimental
spectra, a model for the analysis
of data is required, where the $\Gamma_{2}/\Gamma_{1}$-ratio plays
an important role. This point is further discussed in the next
paragraphs. In the present table, two experimental values are shown:
the one from Outa et al.~\cite{ou05}, whom have used the
approximation $N_{nn}/N_{np} \simeq \Gamma_{n}/\Gamma_{p}$,
where $N_{ij}$ represents the total number of $ij$-pairs
emitted in the $\Lambda$-weak decay (this result is
denoted as preliminary by the author). In this
table it is also reported the value by
Sato et al.~\cite{sa05}, that has been
extracted under the assumption of $\Gamma_{2}/\Gamma_{1}=0.35$ and
obtained from single-proton energy spectra.
These $\Gamma_{n/p}^{exp}$-values are consistent with the above
reported one ($\Gamma_{n/p}^{exp} \sim 0.4 \pm 0.1$).
In this table, it is also observed that the ratio ${\Gamma}_{n/p}$
is roughly unaffected by the renormalization procedure.
\begin{table}[h]
\begin{center}
\caption{The same as Table~\ref{gam12}, but for ${\Gamma}_{n}$,
${\Gamma}_{p}$ and the ratio $\Gamma_{n}/\Gamma_{p}$.}
\vspace{1cm} \label{gamnp}
\begin{tabular}{ccccc}   \hline\hline
$model \; int.$ & $renorm.$ &${\Gamma}_{n}$ &
 ${\Gamma}_{p}$ &
${\Gamma}_{n}/{\Gamma}_{p}$
\\  \hline
$\mbox{Nijimegen} \,$ \cite{na77}   & $no$  & 0.213 & 0.819 & 0.260 \\
$\mbox{NSC97f} \,$ \cite{st99}  & $no$      & 0.155 & 0.660 & 0.235  \\
$\mbox{Nijimegen} \,$ \cite{na77}   & $yes$ & 0.154 & 0.593 &  0.260 \\
$\mbox{NSC97f} \,$ \cite{st99}  & $yes$     & 0.112 & 0.478 & 0.234  \\
$\mbox{experiment} \,$ \cite{ou05}  &  &  &  &  $0.56 \pm 0.12 \pm 0.04$  \\
$\mbox{experiment} \,$ \cite{sa05}  &  &  &  &  $0.60^{+0.11+0.23}_{-0.09-0.21}$  \\
\hline\hline \\
\end{tabular}
\end{center}
\end{table}

Before going on, we give a brief overview of how the values of
$\Gamma_{NM}$ and $\Gamma_{n}/\Gamma_{p}$ are extracted from data.
In first place, the hypernuclear weak decay lifetime $\tau$, is an
observable which is related to the total decay width
$\Gamma_{tot}$, as follows,
\begin{equation}
\label{lifet} \tau = \frac{\hbar}{\Gamma_{tot}},
\end{equation}
where $\Gamma_{tot}=\Gamma_{M} + \Gamma_{NM}$, with $\Gamma_{M}$
being the mesonic decay width. The evaluation of $\Gamma_{M}$ is
less controversial than the non-mesonic decay width, which gives
us some confidence on the experimental value for $\Gamma_{NM}$.

The extraction of the $\Gamma_{n}/\Gamma_{p}$-ratio from data is
much more involved. This is because both $\Gamma_{n}$ and
$\Gamma_{p}$, are primary decays, which implies that they take
place within the nucleus and can not be directly measured. The
magnitudes which can be measured are the number of neutrons (protons)
emitted as a consequence of the $\Lambda$-decay, denoted as
$N_{n}$ ($N_{p}$) or also the number of neutron-neutron (neutron-proton)
pairs, named as $N_{nn}$ ($N_{np}$).
Moreover, these numbers are measured within
certain energy-intervals, which allows us to draw the particle
spectra. There are several ways to connect $N_{n}$ and $N_{p}$
(or $N_{nn}$ and $N_{np}$),
with $\Gamma_{n}$ and $\Gamma_{p}$. One of them is the
INC, which is briefly discussed. The INC is one of the most
sophisticated models to extract this ratio from data. Within this
model, the $N_{n}/N_{p}$-ratio is related to the
$\Gamma_{n}/\Gamma_{p}$-ratio through the following
relation~\cite{ga04},
\begin{equation}
\label{nn/np}
\frac{N_n}{N_p} = \frac{N^{\rm 1Bn}_n \displaystyle
\frac{\Gamma_n}{\Gamma_p}+N^{\rm 1Bp}_n + N^{\rm 2B}_n
\left(1+\frac{\Gamma_n}{\Gamma_p}\right)
\frac{\Gamma_{2}}{\Gamma_1}} {N^{\rm 1Bn}_p \displaystyle
\frac{\Gamma_n}{\Gamma_p}+N^{\rm 1Bp}_p + N^{\rm 2B}_p
\left(1+\frac{\Gamma_n}{\Gamma_p}\right)
\frac{\Gamma_{2}}{\Gamma_1} } \, ,
\end{equation}
with an analogous expression for $N_{nn}/N_{np}$.
The quantities $N^{\rm 1Bi}_k$ and $N^{\rm 2B}_k$ are
numerically evaluated within the INC and are independent of the
weak-vertex. To extract $\Gamma_{n/p}^{exp}$ from this expression
one has to assume a particular value
for the $\Gamma_{2}/\Gamma_{1}$-ratio. For
instance, in~\cite{ba06} the results are:
$\Gamma_{n/p}^{exp}=0.43 \pm 0.10$ for $\Gamma_{2}/\Gamma_{1}=0.26$ and
$\Gamma_{n/p}^{exp}=0.46 \pm 0.09$ for $\Gamma_{2}/\Gamma_{1}=0.$
As mentioned, the INC is one model in the data analysis.
In the work done by Sato et
al.~\cite{sa05}, it is reported $\Gamma_{n/p}^{exp}=
0.60^{+0.11+0.23}_{-0.09-0.21}$ for $\Gamma_{2}/\Gamma_{1}=0.35$,
while $\Gamma_{n/p}^{exp}= 0.87 \pm 0.09 \pm 0.21$ for
$\Gamma_{2}/\Gamma_{1}=0$. These results are obtained by
a Monte-Carlo simulation based on GEANT~\cite{cern}
and the INC from~\cite{ra97},
by fitting single-proton energy spectra and using
the $\Gamma_{n/p}$-ratio as a free parameter.
Let us mention that in~\cite{ba07b} and
\cite{ba07}, a microscopic model for the spectra itself has been
developed, where the primary decays $\Gamma_{n}$ and
$\Gamma_{p}$ are one ingredient within the full calculation.
From this point of view, it is the nucleon emission spectra,
rather than the $\Gamma_{n/p}$-ratio, the magnitude which should
be compared with data.

From these last two paragraphs we have tried to call attention on
the fact that the accurate determination of both $\Gamma_{1}$ and
$\Gamma_{2}$ are equally important. We resume now some of the more
frequent approaches on this subject:
\begin{itemize}

\item \textit{Only $\Gamma_{1}$ is evaluated, while
the phase space for $\Gamma_{2}$ is considered, taking
the $\Gamma_{2}/\Gamma_{1}$-ratio as a free parameter.}
In this case, the dynamics in $\Gamma_{2}$ is not evaluated.
A comparison of both $\Gamma_{NM}$ and
$\Gamma_{n/p}$ with data is questionable as the $\Gamma_{2}$
component is arbitrarily varied to achieve the best match with
data. Up to now, from the experimental point of view, it is not possible to
disentangle the individual magnitudes of $\Gamma_{1}$ and
$\Gamma_{2}$ in $\Gamma_{NM}$. As the magnitude of $\Gamma_{2}$ is
sizable compared with $\Gamma_{1}$, there is no ground to avoid
the explicit evaluation of $\Gamma_{2}$. Note that a not-null $\Gamma_{2}$
implies a correlated ground state, which alters the $\Gamma_{1}$
itself.

\item \textit{Both $\Gamma_{1}$ and $\Gamma_{2}$ are evaluated in
nuclear matter, without renormalization}. In this case, the
problem is the simultaneous reproduction of both
$\Gamma_{NM}^{exp}$ and $\Gamma_{n/p}^{exp}$. If we care about
$\Gamma_{NM}^{exp}$, the wide range of variation in the reported
values for $\Gamma_{1}$ and $\Gamma_{2}$, allows to accomplish
also a good agreement with data for $\Gamma_{NM}$, but with wrong
values for $\Gamma_{1}$ and $\Gamma_{2}$, although the sum is
correct. In this case, a small $\Gamma_{1}$ is compensated by some
spurious intensity added by $\Gamma_{2}$. This mistake is not
harmless as an incorrect ratio $\Gamma_{2}/\Gamma_{1}$ would lead
to a wrong analysis of data and an inappropriate choice for the
transition potential parametrization, which would affect the whole
theoretical calculation. On the other hand, if we focus on
$\Gamma_{n/p}^{exp}$, the theoretical $\Gamma_{NM}$ will certainly
overestimate $\Gamma_{NM}^{exp}$.

\item \textit{The $\Gamma_{1}$ decay is evaluated in finite
nucleus while $\Gamma_{2}$ in evaluated in nuclear matter}. If the
renormalization is not taken into account, the objection rise in
the last paragraph holds here. But the renormalization
procedure is not possible in this case, because in this hybrid model (with
$\Gamma_{1}$ calculated in finite nucleus and $\Gamma_{2}$ in
nuclear matter), there exists no partial decay
width-$\Gamma_{1}(k_{F})$. Let us recall that in the
renormalization procedure, each partial decay widths,
$\Gamma_{1}(k_{F})$ and $\Gamma_{2}(k_{F})$, are multiplied by
the function ${\cal N}(k_{F})$, and then integrated over $k_{F}$.

\item \textit{Both $\Gamma_{1}$ and $\Gamma_{2}$ are evaluated in
finite nucleus.} In addition, a renormalization procedure should
be implemented. The problem here is that due to the huge amount of
possible $3p2h$-configurations, the $\Gamma_{2}$-evaluation is
quite involved. There exists no contribution within this approach
yet.

\end{itemize}

As an additional comment on the second point, in~\cite{ba06} the
$\Gamma_{n/p}$-ratio has been studied with a full microscopic
calculation of both $\Gamma_{1}$ and $\Gamma_{2}$, but with no
renormalization. As the renormalization procedure has very little
influence on the $\Gamma_{2}/\Gamma_{1}$-ratio, this analysis is
still valid. However and as expected, the reported value for
$\Gamma_{NM}$ is rather big.

Finally, two points should be addressed. The first one refers to
the transition potential. The hypernuclei decay is one of the most
important source of information about baryon-baryon
strangeness-changing weak interactions. Therefore there is some
kind of dialectical relation between the transition potential and
the decay widths: there are several parametrization of the
transition potential. Once one parametrization is chosen as the
one which gives the best value for $\Gamma_{NM}$ (for instance),
this let us learn something more about the magnitude of the
coupling constants in the transition potential (note that this kind
of analysis should be corroborated by the agreement
between different nuclear models for the evaluation of $\Gamma_{NM}$).
In this sense, it
is usually argued that the tensor force in the one-pion exchange
potential is too strong. The inclusion of the two-pion
exchange~\cite{ji01}, provides with a strong tensor force whose
sign is opposite to the one-pion one. In~\cite{it02}, a two-pion
couple to a $\rho-$ and $\sigma$-mesons is analyzed. In the
present contribution, we have selected the transition potential as
the one described as Nijimegen. In principle, this choice is particular to
our nuclear model for the evaluation of $\Gamma_{NM}$.
The comparison of $\Gamma_{NM}$ (for
a wide range of hypernucleus), between the just mention potential
and the ones with two-pions, would certainly be of interest.
However, this analysis is beyond the scope of the present
contribution.

The second point which deserves attention is the calculation of
single and double coincidence nucleon spectra. This would be done
also in a self-consistent scheme, using the formalism developed
in~\cite{ba07b} and \cite{ba07}. Note that the INC can be
used in two ways: one has been already discussed and
it refers to the extraction of the $\Gamma_{n/p}^{exp}$-ratio
from the measured spectra. The other one is to start with
the theoretical results for the primary decays (${\Gamma}_{n, \, p, \, nn, \ ...}$)
and then predict a theoretical value for the spectra.
This means that if the theoretical value for the spectra
match exactly with the corresponding data, so does
the  $\Gamma_{n/p}$-ratio. In this spirit,
in~\cite{ba07b} and \cite{ba07} a
microscopic model for the spectra has been presented.
At variance with the INC, the
microscopic model naturally has some quantum-interference
terms not contained in the semi-phenomenological INC model. This
microscopic model is still in its preliminary stages and any
improvement over the simple RPA-model of~\cite{ba07b} is feasible
but difficult. Certainly, a good starting point is the selection
for our model, of a particular parametrization of
the transition potential by means of $\Gamma_{NM}$, which has
been one of the subjects of the present work.

\section{Conclusions}

\label{conclusions}
In the present contribution we have called attention on two simple
but relevant aspects in the evaluation of the nonmesonic weak
decay of a $\Lambda$-hypernucleus. The first point is the former
inappropriate way of including $\Gamma_{2}$ in $\Gamma_{NM}$.
This point should not be underestimated: the plain addition
of $\Gamma_{1}$ plus $\Gamma_{2}$ to obtain $\Gamma_{NM}$, adds
spurious intensity. If there are spurious intensity, a good
$\Gamma_{NM}$-results, imply a distorted
$\Gamma_{2}/\Gamma_{1}$-ratio and then, a distorted
$\Gamma^{exp}_{n/p}$. The
second point, refers to the implementation of $SRC$ within the
kinematical conditions of the $\Lambda$-nonmesonic weak decay.
In addition, $\Gamma_{NM}$ has been evaluated by means of
an already developed nuclear matter model, which
employs the same scheme and interactions for
$\Gamma_{1}$ and $\Gamma_{2}$, but with the two
improvements just mentioned. Among the several parameterizations
of the transition potential, with the one named as Nijimegen~\cite{na77},
it has been achieved an excellent agreement between our result for
the total nonmesonic weak decay width for $^{12}_{\Lambda}C$ and
the corresponding experimental value. This gives us a mechanism to
select the transition potential which is more appropriate for
our model, in view of a forthcoming calculation of the
nucleon emission spectra.

\section*{Acknowledgments}
This work has been partially supported by the CONICET,
under contract PIP 6159.


\begin{thebibliography}{00}

\bibitem{ra98}
E. Oset and A. Ramos, Prog. Part. Nucl. Phys. {\bf 41} (1998) 191.

\bibitem{al02}
W. M. Alberico and G. Garbarino, Phys. Rep. {\bf 369} (2002) 1; in
{\em Hadron Physics}, IOS Press, Amsterdam, 2005, p.~125. Edited
by T. Bressani, A. Filippi and U. Wiedner. Proceedings of the
International School of Physics ``Enrico Fermi", Course CLVIII,
Varenna (Italy), June 22 -- July 2, 2004.

\bibitem {ra97}
A. Ramos, M. J. Vicente-Vacas and E. Oset, Phys. Rev. {\bf C 55}
(1997) 735; {\bf 66} (2002) 039903(E).

\bibitem{ga03}
G. Garbarino, A. Parre\~no and A. Ramos, Phys. Rev. Lett. {\bf 91}
(2003) 112501.

\bibitem{ga04}
G. Garbarino, A. Parre\~no and A. Ramos, Phys. Rev. {\bf C 69}
(2004) 054603.

\bibitem{ba06}
E. Bauer, G. Garbarino, A. Parre\~no and A. Ramos,
nucl-th/0602066.

\bibitem {du96}
J. F. Dubach, G. B. Feldman, B. R. Holstein and L. de la Torre,
Ann. Phys. (N.Y.) {\bf 249} (1996) 146.

\bibitem {os85}
E. Oset and L. L. Salcedo, Nucl. Phys. {\bf A 443} (1985) 704.

\bibitem {ou05}
H. Outa et al., Nucl. Phys. {\bf A 754} (2005) 157c.

\bibitem{Sz91}
J. J. Szymanski et al., Phys. Rev. {\bf C 43} (1991) 849.

\bibitem{No95}
H. Noumi et al., Phys. Rev. {\bf C 52} (1995) 2936.

\bibitem{sa05}
Y. Sato et al., Phys. Rev. {\bf C 71}, 025203 (2005).

\bibitem {ba03}
E. Bauer and F. Krmpoti\'c, Nucl. Phys. {\bf A 717} (2003) 217.

\bibitem{ba04}
E. Bauer and F. Krmpoti\'c, Nucl. Phys. {\bf A 739} (2004) 109.

\bibitem {va92}
D. Van Neck, M. Waroquier, V. Van der Sluys and J. Ryckebusch,
Phys. Lett. {\bf B 274} (1992) 143.

\bibitem {os82}
E. Oset, H. Toki and W. Weise, Phys. Rep. {\bf 83} (1982) 281.

\bibitem {ra94}
A. Ramos, E. Oset and L. L. Salcedo, Phys. Rev. {\bf C 50}, 2314
(1994).

\bibitem{al91}
W. M. Alberico, A. De Pace, M. Ericson and A. Molinari, Phys.
Lett. {\bf B 256}, 134 (1991).

\bibitem {al84}
W. M. Alberico, M. Ericson and A. Molinari,
Ann. Phys. {\bf 154} (1984) 356.

\bibitem {pa97}
A. Parre\~no, A. Ramos and C. Bennhold, Phys. Rev. {\bf C 56}
(1997) 339; \\
A. Parre\~{n}o and A. Ramos, Phys. Rev. {\bf C 65} (2002) 015204.

\bibitem {na77}
M. N. Nagels, T. A. Rijiken and J. J. de Swart,
Phys. Rev. {\bf D 15} (1977) 2547; \\
P. M. M. Maessen, T. A. Rijiken and J. J. de Swart, Phys. Rev.
{\bf C 40} (1989) 2226.

\bibitem {st99}
V. G. J. Stoks and Th. A. Rijken, Phys. Rev. {\bf C 59} (1999)
3009; Th. A. Rijken, V. G. J. Stoks and Y. Yamamoto, {\it ibid.}
{59} (1999) 21.

\bibitem {ma87}
R. Machleidt, K. Holinde and Ch. Elster; Phys. Rep. {\bf 149}
(1987) 1.

\bibitem {br96}
M. B. Barbaro, A. De Pace, T. W. Donnelly and A. Molinari, Nucl.
Phys. {\bf A 596} (1996) 553.

\bibitem{ba07b}
E. Bauer, Nucl. Phys. {\bf A 796} (2007) 11.

\bibitem{cern}
CERN Program Library Entry W5013, $GEANT$.

\bibitem{ba07}
E. Bauer, Nucl. Phys. {\bf A 781} (2007) 424.

\bibitem{ji01}
D. Jido, E. Oset and J. E. Palomar, Nucl. Phys. {\bf A 694} (2001)
525.

\bibitem{it02}
K. Itonaga, T. Ueda and T. Motoba, Phys. Rev. {\bf C 65} (2002)
034617.

\end{thebibliography}
\end{document}